\documentclass[aps,showpacs,nofootinbib]{revtex4}

\usepackage{graphicx}
\usepackage{graphics}
\usepackage{amssymb}
\usepackage{amsmath}
\usepackage{bm}

\newcommand{\be}{\begin{equation}}
\newcommand{\ee}{\end{equation}}
\newcommand{\bea}{\begin{eqnarray}}
\newcommand{\eea}{\end{eqnarray}}
\newcommand{\beal}{\begin{align}}
\newcommand{\eal}{\end{align}}
\newcommand{\bespl}{\begin{split}}
\newcommand{\espl}{\end{split}}

\newcommand{\nslash}{\kern 0.2 em n\kern -0.50em /}
\newcommand{\kslash}{\kern 0.2 em k\kern -0.45em /}
\newcommand{\pslash}{\kern 0.2 em p\kern -0.50em /}
\newcommand{\Sslash}{\kern 0.2 em S\kern -0.50em /}
\newcommand{\Pslash}{\kern 0.2 em P\kern -0.50em /}
\newcommand{\Rslash}{\kern 0.2 em R\kern -0.50em /}

\begin{document}

\title{
Modifications of the 
analytical properties of the two-point correlation function 
in Drell-Yan scattering, caused by rescattering-induced 
decorrelation of the initial hadron state.  
} 

\author{A.~Bianconi}
\email{andrea.bianconi@bs.infn.it}
\affiliation{Dipartimento di Chimica e Fisica per l'Ingegneria e per i 
Materiali, Universit\`a di Brescia, I-25123 Brescia, Italy, and\\
Istituto Nazionale di Fisica Nucleare, Sezione di Pavia, I-27100 Pavia, Italy}

\begin{abstract}

In a recent work, factorization breaking interactions and T-odd 
distributions have been analyzed from the point of view of statistical 
mechanics. One of the central points was the 
process leading from the coherent initial hadron state to 
an incoherent set of short-range parton states, in presence of 
factorization breaking interactions. 
This time-evolution was 
named ``partonization'', or also ``decorrelation'', to distinguish 
it from an ordinary hadron-quark-spectator splitting vertex.  
Here an example is shown for a correlation amplitude 
describing an extreme case for such a process, where the initial 
state phases are lost, but the probability distributions are 
perfectly conserved. 
This requires introducing degrees of freedom that are not normally 
considered in a factorized scheme, and that must disappear at some 
stage of the calculations for the factorization scheme to survive. 
The explicit time-evolution character of the decorrelation process 
disappears when these degrees of freedom are integrated, but the 
integrated correlator presents modified analytical properties 
as a consequence. 
\end{abstract}


\maketitle

\section{Introduction}

\subsection{General}

The study of the so-called 
T-odd distribution functions 
has made it necessary to reconsider some of the basic properties 
of the parton 
model\cite{Feynman} and of the related factorization schemes 
(for the points concerning us here, 
see refs.\cite{CollinsSoperSterman,Bodwin}). 
We may roughly identify two main research lines on T-odd 
distributions. 
 
One starts with the first T-odd ``official'' 
distribution, 
the Sivers function\cite{Sivers} in 1990, followed by the 
Boer-Mulders-Tangerman 
function\cite{MuldersTangerman96,BoerMulders98,Boer99}. 
The former was used to explain single spin 
asymmetries\cite{ABM95}, the latter to explain 
unpolarized Drell-Yan azimuthal 
asymmetries\cite{Boer99}. 

The other research line is 
older\cite{EfremovTeryaev82,QiuSterman91,
BoerMuldersTeryaev97} 
and is centered on the study of 
twist-3 (with respect to $q_T$) effects. It has been very recently 
demonstrated\cite{BoerMuldersPijlman03,JQVY06}, 
that the two things ($q_T-$twist-3 contributions and Sivers function) 
describe the same phenomenon in different $q_T-$regions, and produce 
effective distributions with the 
same $Q^m-$dependence in Drell-Yan 
(the latter describes objects that are twist-3 
with respect to $q_T$, but in Drell-Yan $q_T$ and $Q$ are independent). 

For completeness, we should note that 
single spin asymmetries and unpolarized azimuthal Drell-Yan asymmetry 
have also been interpreted 
via soft mechanisms\cite{BorosLiangMeng}, although still 
according with schemes that 
may be (qualitatively) considered ``T-odd'' for the role played 
by initial state interactions. 

In nuclear physics a time reversal 
odd structure function, the so-called ``fifth structure function'',  
was introduced in 1983 by Donnelly\cite{Donnelly} 
and much later modeled\cite{BB95,BR97} to describe normal 
asymmetries 
in $A(\vec e,\vec e' p)$ quasi-elastic scattering 
(for reviews see \cite{BGPR}). Although the formalism 
presents differences, 
several connections with the partonic T-odd functions   
are present. A point that is exploited later is the 
observation\cite{BR97} that a key passage for producing a nonzero 
fifth structure function is giving absorption features to the 
rescattering terms. 

In hadronic physics, a key role is played by the gauge restoring 
field operator entering the definition of parton 
distribution\cite{CollinsSoperSterman}. In an approach 
where factorization is assumed from the very beginning, this 
field is neglected, but in such case 
leading twist T-odd distribution functions 
are  forbidden\cite{Collins93} by general 
invariance principles. 
After a subtle analysis\cite{BrodskyEtAl02} had shown that 
proper taking into account the gauge factor implies leading  
twist corrections 
in small-$x$ deep inelastic scattering,  
an explicit mechanism was shown to produce a 
nonzero T-odd distribution in a QCD 
framework\cite{BrodskyHwangSchmidt02,BoerBrodskyHwang03}. 
In \cite{Collins02} 
the existence problem was systematically fixed 
by analyzing the time-reversal 
properties of the gauge factor, and in \cite{JiYuan02,BelitskyJiYuan03} 
the pitfalls associated with the limiting behavior 
of the gauge fields have been examined in depth. 
Serious efforts 
has been undertaken\cite{JiMaYuan,BoerMuldersPijlman03,JQVY06} 
to rewrite factorization rules for processes where partonic 
distributions depend on (soft values of) transverse momentum and  
the gauge factor is assumed to play a role. 
In general however, the impossibility to neglect the gauge field 
obliges one to reconsider the correctness, or at least the 
effectiveness, of the 
assumed factorization\cite{BBMP,RatcliffeTeryaev07,
CollinsQiu07,QiuVogelsangYuan07}. 

Several authors have given contributions to the huge development 
of this field in the last few years. 
Models\cite{Yuan03,GambergGoldsteinOganessyan03,
BacchettaSchaeferYang04,LuMa04} 
and studies\cite{Pobylitsa03,DalesioMurgia04,Burkardt04,
Drago05,GoekeMeissnerMetzSchlegel06,BoerVogelsang06,Entropy3} 
around T-odd distribution functions have been 
produced. Recently phenomenological 
parameterizations of the Sivers 
function for quarks\cite{Torino05,VogelsangYuan05,
CollinsGoeke05,BR06a} and gluons\cite{AnselminoDalesioMelisMurgia06} 
have been deduced 
from available 
data\cite{Star,Hermes,Compass,Phenix}, and also 
parameterizations\cite{Boer99,BaroneLuMa05} 
of the Boer-Mulders-Tangerman function have been produced. 
Quite a few experiments aimed at the measurement of T-odd functions 
are planned for the next ten 
years\cite{panda,assia,pax,rhic2,compassDY}. 

\subsection{Starting hypothesis and aims of this work} 

In a previous work a picture was presented of T-odd functions as a 
result of a statistical process associated with initial state 
interactions between two hadrons. This transformed the hadron structure,  
that initially was characterized by long-range correlations 
(``hadronic phase'') 
into an incoherent collection of 
short-range objects describing a quark in the quark-parton model 
(``partonic phase''). So, the presence of a nonzero T-odd 
function was related to the decay properties of the initial 
state wavefunction, and to a singular rate of increase of the 
entropy of the system. 

A point that was not discussed in that work was the way this 
decorrelation 
process could show information about itself in the two-point 
amplitude describing a semi-inclusive measurement, in 
particular a Drell-Yan measurement (the so-called correlator of the 
process, see next section). The presence of such information 
in the correlator is essential to have a connection 
between the physical phenomenon (the 
decorrelation process) and the possibility to observe it 
in the distribution functions. 

The main section of this work is section V, where 
we show an  
example of a correlation amplitude describing 
a partonization/decorrelation process. This two-point amplitude 
contains in itself an initial pure state, a final incoherent 
probability distribution, and the transition between the two 
regimes. 

A peculiarity of the decorrelation process 
is brought to its extreme by the chosen correlator: 
the hadron-to-parton transition is ``ideal'', i.e. although 
phases are destroyed in the transition, probabilities are perfectly 
conserved. 
We will speak of ``ideal decorrelation/partonization process''.

The description of a process involving factorization-breaking 
interactions 
(causing the decorrelation) requires introducing extra degrees of 
freedom 
in addition to those associated with $x$ and $\vec k_T$, the 
quark longitudinal fraction and transverse momentum. There is nothing 
exotic in this. Any model considering rescattering introduces 
such extra coordinates, and several of the above quoted references do 
this. In the following we will name ``extra'' degrees of freedom 
those that are not present in a standard factorization regime. 

Let $\xi$ $\equiv$ $P_+z_-$  
be the spacetime coordinate conjugated with the quark 
longitudinal fraction $x$ (see section II for all the definitions), 
and let $G(\xi)$ be the correlation amplitude 
in a factorization-respecting approach.  
Let 
$\eta$ be an extra coordinate introduced by  
rescattering. 
A $G(\xi)$ correlator that at least formally 
respects factorization must originate in an integration 
process of the form $G(\xi)$ $=$ $\int d\eta G'(\xi,\eta)$. 

We will consider a decorrelation process 
that is evident in the pre-integration 
amplitude $G'(\xi,\eta)$, but not in $G(\xi)$. 
We will show that $G(\xi)$ still keeps in 
itself a track of the decorrelation process, that a hard probe 
is able to detect in the form of a T-odd 
additional distribution function. This track has to do with the 
analytical properties of the correlator. 

Distribution functions are calculated from the imaginary part of 
the correlator. This imaginary part derives from a real particle 
production cut, and the cut position has an infinitesimal imaginary 
shift $i 0$ from the real axis 
associated with causality. We will show that the imaginary 
shift becomes $finite$: $i 0$ $\rightarrow$ $i n$, $n$ $>$ 0, 
when the proposed correlator is averaged over extra degrees of 
freedom. 

In another work\cite{ShortLetter1} it is shown that such a finite 
shift may imply T-odd observable corrections to the 
distribution functions. Here we only discuss the formation of the 
finite imaginary shift. 

We stress that 
in the following nothing is demonstrated about factorization. But 
we require the assumption that factorization breaking is a small
effect, because this allows for simplifying approximations. 
In concrete terms, 
``small effect'' means that the parameter $n$ (appearing in the 
following associated with rescattering corrections to the 
correlator) is small. 

\subsection{Additional degrees of freedom} 

The relevant distribution functions are extracted from a 
two-point correlation function $G(1,2)$. This  
describes the propagation of a quark hole 
created in a hadron in the spacetime position ``1'' to the 
spacetime position ``2''. 

In presence of factorization, $G$ depends on ``1'' and ``2'' 
via the difference ``1--2'' only (the absence of rescattering 
implies translation invariance). In addition, one of the two light-cone 
projections of ``1--2'' must play no role. 

There are several degrees of freedom that one may imagine 
to acquire a role when factorization breaking interactions are 
included. First, the ones related with ``1+2'', or with 
the neglected light-cone component, or with the position and 
internal structure of the target hadron. 

To simplify things, we limit here to $one$ extra degree of freedom.  
This is sufficient to understand how decorrelation develops 
and how it leads to dissipative effects when averaged. 
We neglect $z_+$ and the other possible variables, 
but include ``1+2'' (specifically: 
$z_-(1)+z_-(2)$) in our calculations.  

\section{General definitions and notations} 

We consider a Drell-Yan process taking place between two colliding 
hadrons with momenta $P_\mu$ and $P_\mu'$, in their center of mass 
frame. 
The parton we will examine in the 
following is a quark belonging to the former hadron. 

To select leading twist terms, we will use the infinite 
momentum limit $P_+$ $\rightarrow$ $\infty$. In this limit 
$(P + P')^2$ $\sim$ $P_+P_-'$ 
$\sim$ ${P_+}^2$ 
is infinite, and for finite $x$ and $\bar{x}$  
also $Q^2$ $\equiv$ $x\bar{x} (P + P')^2$ is infinite. 

The relevant amplitude 
considered in this work is the traditional one 
where a hole is created in a hadron state 
by extracting a quark/antiquark in a point $z_\mu'$, the 
``hadron+hole'' set propagates across the cut through real states 
only, and in $z_\mu''$ the hole is filled and the original hadron 
state is restored 
by reinserting the quark in its place. All the 
observable quantities interesting us are extracted from the imaginary 
part of this amplitude, that in this paper we simple 
name $G(1,2)$. Where necessary, the spacetime coordinates 
corresponding to ``1'' and ``2'' are explicitly 
written. 
Variables corresponding to ``1'' are indicated as 
$z'$, $k'$ etc, variables corresponding to ``2'' with 
$z''$, $k''$ etc. 

The extracted quark has momentum 
$k_\mu$ with lightcone 
$k_+$ $\equiv$ $x P_+$ and negligible $k_-$. The impact 
parameter $\vec b$ is the space vector conjugate to $\vec k_T$. 
The spacetime displacement of the hole from $z_\mu'$ to $z_\mu''$ 
is $z_\mu$ $\equiv$ $z_\mu'-z_\mu''$ 
$\equiv$ $(z_-,z_+,\vec b)$ $\approx$ $(z_-,0,\vec b)$. 
The fourth coordinate $z_+$ plays no role and 
is not explicitly reported in the following. 

Any relevant distribution function $q(x)$ is the 
Fourier transform 
with respect to $z_-$ and $\vec b$ of a correlation function 
$g(z_-,\vec b)$. 
Leading twist effects are naturally selected by writing the 
Fourier transform with respect to the scaled variables 
$x$ and $\xi$: 

\begin{equation}
\xi\ \equiv\ P_+ z_-, \label{eq:def_xi}
\end{equation}

\begin{equation}
q(x,\vec k_T)\ \equiv\ P_+ \int e^{-i xP_+z_-} e^{ik_T b}
g(z_-,\vec b) dz_- d \vec b\ \equiv\ 
\int e^{-i x\xi} e^{ik_T b}
G(\xi,\vec b) d\xi d\vec b. \label{eq:def_qx}
\end{equation}

$G(\xi,\vec b)$, $\xi$, $x$, etc will be named ``scaled'' quantities. 
The range of useful values of $\xi$ remains 
finite $\sim$ $1/x$ when $P_+$ $\rightarrow$ $\infty$. 
In this limit 
the function $G(\xi)$ does 
not need to be singular in the origin to produce a nonzero $q(x)$.  
So, in the following the discussion is in terms of the 
scaled quantities. 

$G$ is here used as a function and not a matrix,  
so it must be read as a 
projection of the spinor correlation matrix $G_{ij}$ over a suitable 
operator, e.g. $\gamma^+$. The $\gamma^+$ projection 
selects simultaneously the main 
unpolarized 
quark distribution and the Sivers T-odd function. 

The two functions $q_(x,\vec k_T)$ and $G(\xi,\vec b)$ are related by 
double Fourier transformation with respect to $x/\xi$ and 
$\vec k_T/\vec b$. In the following, 
we will mostly work on the mixed representation function $f(x,\vec b)$: 

\begin{equation}
f(x, \vec b)\ \equiv\ 
\int e^{-i x\xi} 
G(\xi,\vec b) d\xi. \label{eq:def_fx}
\end{equation}
$G$ and $f$ depend on longitudinal and transverse variables. 

Where not necessary to resolve ambiguity, 
the transverse variables will $not$ be explicitly written, since 
the amount of 
nonstandard variables and constants 
$\eta$, $\Delta$, etc needed in the following is rather large. 
So, e.g. the above eq.\ref{eq:def_fx} may simply assume the form 

\begin{equation}
f(x)\ \equiv\ 
\int e^{-i x\xi} 
G(\xi) d\xi \label{eq:def_fx2}
\end{equation}
where the presence of $\vec b$ is implicitly assumed on both sides. 
This should cause no ambiguity, since in this paper $\vec k_T-$integrated 
quantities do not appear: all the considered functions depend on 
both longitudinal and transverse variables. 


In the following, we will have to extract $G(\xi)$ $\equiv$ 
$G(\xi'-\xi'')$ from the 
integration of a 
two-point amplitude $G(\xi',\xi'')$ over $\xi'+\xi''$. 
The latter 
can be written in terms of two independent variables $x'$ and $x''$ via 
a double Fourier transform $exp[ -i (x'\xi' - x''\xi'') ]$. 
To separate degrees of freedom, we use 
the definitions 

\begin{equation}
\xi\ \equiv\ \xi'-\xi'',\hspace{0.5truecm} 
\eta\ \equiv\ {{\xi'+\xi''} \over 2}.
\label{eq:def5}
\end{equation}

\begin{equation}
X\ \equiv\ {{x'+x''} \over 2},\hspace{0.5truecm} \Delta\ =\ x'-x'', 
\label{eq:def6}
\end{equation}

then we have 
\begin{equation}
exp[-i (x'\xi' - x''\xi'')]\ =\ exp[-i ( X\xi + \Delta \eta )]
\label{eq:def7}
\end{equation}

\section{Light-cone translation averaging operator} 

In presence of a violation of translation invariance, 
we could define $f_A(x)$ as an average over $\eta$ $=$ $(\xi'+\xi'')/2$: 

\begin{equation}
f_A(x)\ \equiv\ {1 \over A} \int_0^A d\eta 
\int e^{-i x\xi} G(\xi',\xi'') d\xi ,
\label{eq:average1}
\end{equation} 

\noindent 
and a distribution 
$f(x)$ could be defined as 
$f_A(x)$, if $A$ is large enough to wash away the effects of 
$\eta-$irregularities. 
To simplify notations we define\footnote{The integration 
range may be also taken as symmetric from $-A$ to $+A$ 
(from $-\infty$ to $+\infty$ below in the definition of $K_M$). 
Indeed, we have $\xi'$ $>$ $\xi''$ due to intermediate real state 
propagation from $\xi''$ to $\xi'$, but this does not touch 
$\eta$ $=$ $(\xi'+\xi'')/2$. Since all the relations contained 
in this work are even with respect to $\eta$, one may limit 
to positive $\eta$ only. 
} 

\begin{equation}
[\hat I_A G](\xi)\ \equiv\ 
{1 \over A} \int_0^A G(\xi',\xi'') d\eta
\label{eq:ia}
\end{equation}

\noindent 
assuming that the dependence on $\eta$ is completely lost 
when the average covers a large enough range $A$. 
We may rewrite the above eq.\ref{eq:average1}: 
\begin{equation}
f_A(x)\ \equiv\ 
\hat I_A \int e^{-i x\xi} G(\xi',\xi'') d\xi\  
=\ 
\int d\xi e^{-i x \xi} \big[\hat I_A G\big](\xi). 
\label{eq:ia2}
\end{equation}

\noindent
Since the $\xi'+\xi''$ and $\xi'-\xi''$ integrations can always 
be exchanged as we have just done, 
a translation invariant object may always be defined, for 
a large enough $A$. 

The factor $1/A$ is not necessary but guarantees a correct 
normalization of final results, according with later equations 
in this work. One may equivalently 
define $\hat I_A$ as a $sum$ operator, rather than an average 
operator. However, either in $\hat I_A$, or in the definition 
of the wavefunctions entering the correlation function, a 
normalization must be present. Formally, $\hat I_A$ may be 
applied, redundantly,  
also in absence of factorization breaking effects. Then, the 
normalization $1/A$ leaves 
the results unchanged, without the need of introducing further 
normalization factors in the wavefunctions. 

Instead of $\hat I_A$, we could use another averaging operator:  

\begin{equation}
[\hat K_{M}G](\xi)\ \equiv\ {1 \over N} \int_{-M}^{M} ds 
\int_{0}^{+\infty} d\eta e^{-i\eta s} G(\xi',\xi'')
\label{eq:kmn}
\end{equation}

\noindent 
where $s$ is the longitudinal fraction 
conjugated with $\eta$. 
This may look more appropriate than $\hat I_A$, since it 
introduces the 
double Fourier transform eq.\ref{eq:def7}, and the sum is over 
a momentum-like loop variable. 
However, $\hat K_{M}$ is 
equivalent to $\hat I_A$, for $N$ $=$ $A$ and $M$ $=$ $\pi/A$. 
Exchanging the $\eta$ and $s$ integrations,  and applying 
the operator to 
$G(\eta,\xi)$ $\equiv$ $G(\xi,\xi'')$, 

\begin{equation}
\int_0^\infty d\eta G(\eta,\xi) 
\int_{-M}^{M} ds e^{-i\eta s}\ =\ 
2 \int_0^\infty d\eta G(\eta,\xi) 
{{sin(M\eta)} \over \eta}\ 
\approx\ 
2 \int_{0}^{\pi/M} d\eta G(\eta,\xi).  
\end{equation}

So from now on we will use directly $\hat I_A$. 

\section{Enlarged 
delta-functions}

If the problem is $\eta-$independent, 
a $\delta(x'-x'')$ function associated 
with the conservation of the light-cone momentum is present 
(according with eqs. \ref{eq:def6} 
and \ref{eq:def7}, $x'-x''$ is the variable conjugated with $\eta$). 
In presence of interactions that violate 
$x-$conservation with $\vert x'-x''\vert$ $\sim$ $n$, 
we need substituting $\delta(x'-x'')$ with a function that 
takes this nonconservation into account. 

Exploiting a technique developed in \cite{BR97} and \cite{BR_JPG} 
to roughly but quickly approximate the resummed effect of rescattering, 
we define an ``enlarged $\delta-$function'' 
$\delta_n(x'-x'')$: 

(i) $x'$ $\approx$ $x''$ within gaussian uncertainty 
$\vert n \vert$, 

(ii) $\int_0^1 dx'' \delta_n(x'-x'')$ $\approx$ 1 (with exact equality 
for an infinite integration range). 

Such functions are 
normally used to define the ordinary $\delta(x'-x'')$ 
function
via their 
$n$ $\rightarrow$ 0 limit. 

Several equivalent representations for $\delta_n(x'-x'')$ 
are possible, and in the following we will use different ones 
interchangeably, accordingly with mathematical convenience. 
Also, factors like $\sqrt{3}$ or 
$\pi$ will be normally neglected. 

A useful representation is\footnote{From now on we 
use $\Delta$ $\equiv$ $x'-x''$, $X$ $\equiv$ $(x'+x'')/2$, see 
eq.\ref{eq:def6}.} 

\begin{equation}
\delta_{n}
(\Delta)\ \equiv\ 
{1\over \pi} {n \over {n^2 + \Delta^2}} 
\nonumber
\end{equation}
\begin{equation}
\equiv\ \delta_{n+}(\Delta)\ 
+\ \delta_{n-}(\Delta)
\equiv\ 
{i\over {2\pi}} {1 \over {\Delta + in}}\ -\ 
{i\over {2\pi}} {1 \over {\Delta - in}}. 
\end{equation} 

Causality implies the substitution 
\begin{equation}
\delta_{n}(\Delta)\ \rightarrow 
\delta_{n+}(\Delta). 
\end{equation}
that corresponds, in its Fourier transform, to 
\begin{equation}
exp(-n \vert \xi \vert)\ \rightarrow 
\theta(\xi)exp(- n \xi) 
\end{equation} 

Since $\xi$ $>$ 0 in the relations interesting us, in the following 
$\delta_n$ must be meant as $\delta_{n+}$, i.e. 
one pole only is considered in the complex plane integrations. 
At the condition that $n$ is small enough to justify the 
idea that the (real-axis) integral is dominated by the peak 
at $\Delta$ $=$ 0, we may approximate the integral 
with the contribution from the pole $x'$ $=$ $x+in$ of the 
function $\delta_{n+}$ (see \cite{ShortLetter1} for 
a more precise justification of this): 

\begin{equation}
\int_0^1 dx' f(x') \delta_{n}(x-x')\ \approx\ f(x + in)\ 
\approx\ f(x)\ + {i n} {{df} \over {dx}}\ +\ O(n^2). 
\label{eq:approx1}
\end{equation}

In the following, we need the 
relation\footnote{The use of the sign ``$\approx$'' instead of ``$=$'' 
is due to (i) a constant factor of magnitude unity, (ii) two 
different representations are used for the functions 
$\delta_n$ appearing in the left and right hand sides of 
the equation, 
and the corresponding functions are equal (in distribution 
sense) only in the 
$n$ $\rightarrow$ 0 limit.
}

\begin{equation}
\hat I_A [\delta_{\gamma/\eta}(\Delta)]\ \approx\  
\delta_{\gamma/A}(\Delta)\ 
\label{eq:delta_abgamma} 
\end{equation}

To show its validity, we use the another representation for an 
enlarged $\delta-$function: 

\begin{equation}
\delta_{\gamma/\eta}(\Delta)\ \equiv\ {1 \over {2\pi}} 
\int_{-\vert \eta \vert /\gamma}^{+ \vert \eta \vert /\gamma} 
e^{iy\Delta} dy\ 
\end{equation}
When 
$\hat I_A$ is applied to it, we get 

\begin{equation}
2\pi\hat{\delta_n}(\Delta) \equiv\ 
\hat I_A[\delta_{\gamma/\eta}(\Delta)]\ =\ 
{1 \over A} 
\int_0^A d\eta \int_{-\vert\eta\vert/\gamma}^{+\vert\eta\vert/\gamma} 
e^{i\eta'\Delta} d\eta' 
\end{equation}
where we have defined 
\begin{equation}
n\ \equiv\ \gamma/A. 
\end{equation} 

That $\hat \delta_n$ is really an enlarged$-\delta$ function may be 
demonstrated by direct 
integration of the previous equation:  
\begin{equation}
\hat \delta_n(\Delta)\ =\ {1 \over {2\pi n}} 
{{sin^2(\Delta/n)} \over {(\Delta/n)}^2}. 
\end{equation}

The right hand side function has $n-$independent $\Delta-$integral $= 1$  
and limiting behavior (for small $n$) like a delta function (its 
value is $\infty$ for $\Delta$ $=$ 0 and 0 for any other $\Delta$). 
So it is a fair representation for an $n$-enlarged $\delta$ 
with precise width $\sqrt{3}n$ 
(from the small-$\Delta$ power series 
for it). 


\section{A decorrelating configuration} 

We assume that initial state interactions, and consequently 
the correlator, depend on both 
$\xi$ and $\eta$. 
We introduce a parameter $\gamma$ so that  
the effect of rescattering in terms of $x-$nonlocality decreases at 
increasing values of $\eta$, to loss any effectiveness for 
$\eta$ $>>$ $\gamma$. This makes sense, since a large $\eta$ 
means large values of $\xi'$, $\xi''$ or both (their difference 
$\xi$ may remain small) meaning that the target hadron is far 
and interaction is suppressed. 


A function of $x$ that depends parametrically on 
$\eta$ can be used to model 
an $\eta-$evolution towards decorrelation. 

Let us use the limiting and normalization properties of 
$\delta_{\gamma/\eta}(x'-x'')$: 

\begin{equation}
\delta_{\gamma/\eta}(x-x')\ 
\approx\ \delta(x-x'), \hspace{0.5truecm} for\ \eta\ >>\ \gamma;
\end{equation}
\begin{equation}
\delta_{\gamma/\eta}(x-x')\ \approx\ const, 
\hspace{0.5truecm} for\ \vert x-x'\ \vert << \gamma/\eta,
\end{equation}
\begin{equation} 
\int dx' \delta_{\gamma/\eta}(x-x')\ \approx\ 1. 
\end{equation}

We define

\begin{equation}
\Psi(\xi)\ \equiv\ \int dx \psi(x) e^{i x \xi}, 
\end{equation} 

\begin{equation}
G(\xi',\xi'')\ \equiv\ G(\eta,\xi)\ \equiv\ 
\int dx' dx'' \psi(x')\psi^*(x'') 
e^{i(x'+x'')(\xi'-\xi'')/2} 
\delta_{\gamma/\eta}(x-x'). 
\end{equation}
\begin{equation}
=\ \int dX d\Delta\hspace{0.1truecm} \psi(X + \Delta/2) \psi^*(X - \Delta/2) 
e^{iX\xi} 
\delta_{\gamma/\eta}(\Delta), 
\label{eq:def_G}
\end{equation}

$G$ has the following limiting properties: 

\begin{equation}
G(\xi',\xi'')\ \approx\ 
G_0(\xi), \hspace{0.5truecm} for\ \eta\ <<\ \gamma;
\end{equation}
\begin{equation}
G(\xi',\xi'')\ \approx\ G_\infty(\xi), \hspace{0.5truecm} 
for\ \eta\ >>\ \gamma.
\end{equation}

where 

\begin{equation}
G_0(\xi)\ \equiv\ 
\Psi(\xi/2)\Psi^*(-\xi/2), 
\end{equation}
\begin{equation}
G_\infty(\xi)\ \equiv\ 
\int dX \vert \psi(X) \vert^2  e^{iX\xi}. 
\end{equation}

The former object is the 
correlator of a pure state, 
while the latter is the correlator of an incoherent overlap of 
plane waves $exp(-ix\xi)$. 
Remarkably, in both regimes the probability for extracting a given 
plane wave $exp(ix\xi)$ from the system is the same: 
$\vert \psi(x) \vert^2$. 

So we get an ``ideal partonization process'' with respect to $x$. 
The shrinking width-parameter of the 
enlarged delta handles the transformation from 
the density matrix of a pure state to the one of a probabilistic 
distribution. 

The separation parameter of the two regimes is $\gamma$. 
For $\eta$ much smaller or larger than $\gamma$ we are in one 
of the two regimes. As far as $G$ is not averaged with respect to 
$\eta$, it contains explicitly both regimes. 
Now we need to average $G$ over $\eta$, so to define an object 
that may fit in the factorization formalism. 

We define 
(cfr eqs.\ref{eq:def_fx} and \ref{eq:ia2})

\begin{equation}
f(x)\ \equiv\ \hat I_A \int d\xi e^{-ix\xi} G(\eta,\xi)\ 
=\ \int d\xi e^{-ix\xi} \hat I_A G(\eta,\xi). 
\label{eq:def10}
\end{equation}

$\hat I_A$ performs the average over an $\eta-$range of size $A$. If 
$A$ is large enough to make the coherence region $\eta$ $\sim$ $\gamma$ 
irrelevant, the above eq.\ref{eq:def10} 
is equivalent to the ordinary definition 
of a distribution function, with $f(x)$ $=$ $\vert \psi(x)\vert^2$. 

We assume here $n$ $\equiv$ $\gamma/A$ $<<$ 1, but 
not by several orders. So, we expect the coherence region 
$\eta$ $\sim$ $\gamma$ to have influence, although small, on the 
final average result. 
For small $n$ it is possible to 
use the approximations of section IV. Defining 

\begin{equation}
f_o(x)\ =\ f(x)\vert_{n=0}
\end{equation}
we get (see eq.\ref{eq:approx1}) 

\begin{equation}
f(x)\ \approx\ f_o(x+in)\ \approx\ 
f_o(x) + i n {{d f_o} \over {dx}}\ =\ 
\vert \psi(x)\vert^2 + {i n} 
{{d \vert \psi(x)\vert^2} \over {dx} } 
\label{eq:final1}
\end{equation}

The above approximate equality derives from two 
calculation steps in eq.\ref{eq:approx1}. For the present work 
the interesting one is in the former one: 

\begin{equation}
f(x)\ \approx\ f_o(x+in)
\label{eq:final1b}
\end{equation}

It requires $n$ $<<$ 1 and 
$x$ far from 0 and 1 by an amount of size $n$ at least. 
Its validity is demonstrated in \cite{ShortLetter1} via the 
mapping $z$ $\equiv$ $log[x/(1-x)]$. However it is 
reasonable at intuition level: if the above conditions on $n$ and $x$ 
are respected, one may guess that all the relevant things 
take place in a restricted area of the complex plane near the  
real axis range $(n,1-n)$, and assume that the 
integral is dominated by singularities in this region. 

Eq.\ref{eq:final1b} says that 
a finite width $n$ in $\delta_n(x-x')$ has the effect of 
introducing a finite imaginary part $n$ in the argument of $f_o(x+in)$. 
An infinitesimal imaginary part is present in $f$ 
because of causality, also in no-rescattering regime: 
in the fourier transform $f(x)$ $=$ 
$\int d\xi exp(-ix\xi)G(\xi)$ the lower integration limit is 
zero and not $-\infty$. Since this means that we fourier-transform 
a function of the form $g(\xi)\theta(\xi)$, $f(x)$ actually 
means $f(x+i0)$ (i.e. $\theta(\xi)$ is accompanied by an 
infinitesimal damping that takes care of the potentially dangerous limit 
$\xi$ $\rightarrow$ $\infty$). A finite $n$ makes the 
imaginary shift finite: $i0$ $\rightarrow$ $in$. 
It means $\theta(\xi)$ $\rightarrow$ 
$\theta(\xi)exp(-n\xi)$. 

To assume that $n$ is small allows for the easy 
approximation eq.\ref{eq:final1b}, but the described property 
(the imaginary shift becoming finite) is likely to be more general. 
At intuition level, it means that a diagram where some degrees of 
freedom associated to rescattering 
are averaged away acquires absorption properties, and this is not 
surprising (see the discussion in the next section). 

For small $n$, 
the second step of eq.\ref{eq:final1} allows one to 
write the overall result as a simply readable sum, where one of the 
two terms is not affected by rescattering. 
As shown in \cite{ShortLetter1}, the imaginary correction 
may lead to an observable asymmetry, if it affects 
even-odd interference terms in the partial wave decomposition of 
the two-point amplitude. This allows for the possibility to observe 
the effects of the discussed finite imaginary shift. 

\section{Discussion.}

\subsection{Non-hermitean hamiltonians and hidden degrees of freedom. 
Analogies with Feshbach's theory and the role of 
Fok-Krilov's theorem.} 

An interesting point regards the analogies between the 
discussed argument and some known problems of nuclear and 
statistical physics. 

A nonzero T-odd observable means 
that an effective and not-hermitean hamiltonian is 
operating. ``Effective'', because we know that strong and 
electromagnetic interactions do not violate T-reversal 
invariance. So a really fundamental hamiltonian cannot produce 
T-odd amplitudes. 
On the other side, if our model of the process contains 
a hamiltonian operator that does not really take into account 
all the relevant degrees of freedom, this hamiltonian becomes 
``effectively'' non-hermitean. 

This problem has been long studied 
in nuclear physics (see in particular \cite{Feshbach}). In general 
terms, following the standard 
Feshbach's scheme\cite{Feshbach}, one may write a time-evolution 
equation for the system wavefunction $\psi$, and split the 
space of all the possible solutions (satisfying scattering 
boundary conditions) into ``globally elastic'' states $\psi_E$ 
and ``globally inelastic'' $\psi_I$ 
states.\footnote{The 
word ``global'' 
means that we only require $E(1)$ $=$ $E(2)$, not caring 
what happens in intermediate 
states.} 

The evolution 
equation may be transformed into a system of 
two coupled equations for $\psi_E$ and 
$\psi_I$. Using one of the two equations for writing $\psi_I$ in terms 
of $\psi_E$, one may insert this relation into the other one 
and derive an equation for $\psi_E$ only. The evolution 
of $\psi_E$ is now determined by an effective hamiltonian 
operator that is partly non hermitean. The non-hermitean part 
is determined by ``locally inelastic'' processes, i.e. second order 
processes where an inelastic channel is 
excited as an intermediate state, to finally de-excite into 
the elastic channel. 
The Green function for the ``reduced'' problem 
(the problem for the globally elastic channel only) 
has a pole with a finite imaginary part for the self-energy 
correction: $1/(E-E'+i0)$ $\rightarrow$ $1/(E-E'+i\Gamma)$. 

As well known in nuclear physics, these arguments give a solid 
formal framework for effective, non-hermitean, hamiltonian 
operators (optical potentials, etc), but are rarely useful for 
their direct calculation from first principles. 
However, they demonstrate that hiding degrees of freedom 
leads to an imaginary part in the energy-shell-projected  
amplitude. 

The factorization formalism in hard scattering imposes 
an $x-$conservation shell for the propagating quark hole. 
In a situation where $x$ is not conserved at all (i.e. 
in presence of rescattering) this creates the same situation 
as in the optical potential theory: a projection on a subspace  
of $x-$conserving states  
in one case, of energy-conserving states in the other 
case. 

In the hard scattering case the hamiltonian 
is substituted by a light-cone operator, 
but the general arguments could be repeated step by step, 
just after changing the name of the variables. 

The physical origin of the imaginary component of the potential,  
and consequently of the finite imaginary contribution to the 
self-energy, is interesting. 
It derives from the presence of 
several incoherent intermediate inelastic channels. If we consider 
a wavepacket with $x-$spread $\delta x$, all the single plane wave 
components of this wavepacket are elastically scattered, but 
coherence between them is lost, and the wavepacket shape 
decays in a ``time'' $1/\delta x$. 
This can be considered 
an application of the Fok-Krilov theorem\cite{Krilov}, stating 
that a wavepacket whose frequency spectrum is made 
continuous by an interaction for $t$ $>$ 0 describes an 
irreversible process, and has a decaying self-correlation 
function (see \cite{Entropy3} for extensive discussion about 
this). 

\subsection{Partonization/Decorrelation process within known models} 

In the example of section V 
we have considered an ideal 
limit for the chaotic hadron-to-parton evolution, 
that could be named ``ideal partonization 
process'', where the probabilistic features of the initial 
quark state with respect to $x$   
are perfectly conserved, although phase features 
are lost. 

In terms of known quark-hadron diagrams, 
an example of a quark state in a hadronic phase is 
the hadron-quark-spectator 
vertex employed in the spectator model by 
\cite{JakobMuldersRodriguez97}. Such a splitting diagram describes 
a quantum fluctuation of a hadron into a quark+spectator pair. 
This fluctuation is perfectly reversible to the original 
hadron state. 

The presence of a soft cutoff $\Lambda$ $\sim$ 
$R_{hadron}$ 
in the vertex form 
factor guarantees that no momentum component $k$ $>$ $\Lambda$ 
crosses the vertex, so that correlation is present up to a distance 
$\delta z$ $\sim$ $R$. 

When the splitting takes place  
under the action of initial state interaction with another hadron, 
the process is incoherent and presents a certain degree of 
irreversibility. Then we may speak of ``partonic phase''. 

A typical example of such a process is a diagram 
for the calculation of the Sivers function, 
where a gluon exchange is added to the previous splitting process 
(it may be found in several of the above quoted models, starting 
from \cite{BrodskyHwangSchmidt02}). The added gluon loop 
involves momenta up to the hard scale, and this destroys 
soft states. 

At the lowest order, an example of transition between the two 
situations is a cut diagram where the previous two examples 
interfere. This interference graph is the basic ingredient of 
several calculations of the Sivers function, and literally 
reproduces the light-cone evolution from one of the two phases to 
the other one (taking the causality factor $\theta(\xi)$ into 
account).  

\section{Conclusions} 

In this work 
an example has been shown for the correlator of 
a hadron state, that presents time-evolution properties. 
It initially consists of a pure state, and finally of 
a completely incoherent probabilistic mixture. The 
$x-$distribution is the same in both cases. This evolution 
is caused by rescattering, and is visible 
as far as the additional degrees of freedom associated with 
rescattering 
are explicitly visible. After these degrees of freedom have 
been integrated, the remaining correlator shows modified 
analytical properties, consisting in a finite imaginary 
shift of the main cut singularity of the process.  




\begin{thebibliography}{99}


\bibitem{Feynman}
R.P.Feynman, ``Photon-Hadron interactions'' (Benjamin, Reading, MA, 1972). 

\bibitem{CollinsSoperSterman}
J.C.Collins, D.E.Soper and G.Sterman, 
Nucl.Phys. {\bf B261} 104 (1985); 
J.C.Collins, D.E.Soper and G.Sterman, 
Nucl.Phys. {\bf B308} 833 (19888). 

\bibitem{Bodwin}
G.Bodwin, Phys.Rev. {\bf D31} 2616 (1985); 
G.Bodwin, Phys.Rev. {\bf D34} 3932 (1986); 



\bibitem{Sivers}
D.Sivers, Phys.Rev. {\bf D41}, 83 (1990); 
D.Sivers, Phys.Rev. {\bf D43}, 261 (1991).

%

\bibitem{MuldersTangerman96}
P.J.Mulders and R.D. Tangerman, Nucl.Phys.{\bf B461}, 197 (1996); 

\bibitem{BoerMulders98}
D.Boer and P.J.Mulders, Phys.Rev. {\bf D57}, 5780 (1998); 

\bibitem{Boer99}
D.Boer, Phys.Rev. {\bf D60}, 014012 (1999); 

\bibitem{ABM95}
M.Anselmino, M.Boglione and F.Murgia, 
Phys.Lett.{\bf B362}, 164 (1995); 

\bibitem{EfremovTeryaev82}
A.V.Efremov and O.V.Teryaev, Sov.J.Nucl.Phys. 
{\bf 36} 140 (1982) [Yad. Fiz. {\bf 36} 242 (1982)]; 
A.V.Efremov and O.V.Teryaev, 
Phys.Lett.{\bf B 150} 383 (1985). 

\bibitem{QiuSterman91}
J.Qiu and G.Sterman, 
Phys.Rev.Lett.{\bf 67} 2264 (1991); 
J.Qiu and G.Sterman, 
Nucl.Phys.{\bf B 378} 52 (1992); 
J.Qiu and G.Sterman, 
Phys.Rev.{\bf D 59} 014004 (1999). 

\bibitem{BoerMuldersTeryaev97}
D.Boer, P.J.Mulders and O.V.Teryaev, 
Phys.Rev.{\bf D 57}, 3057 (1997). 

\bibitem{BoerMuldersPijlman03}
D.Boer, P.J.Mulders and F. Pijlman, 
Nucl.Phys. {\bf B 667}, 201 (2003).


\bibitem{JQVY06}
X.Ji, J.W.Qiu, W.Vogelsang and F.Yuan, 
Phys.Rev.Lett.{\bf 97} 082002 (2006); 
X.Ji, J.W.Qiu, W.Vogelsang and F.Yuan, 
Phys.Rev.{\bf D 73}, 094017 (2006). 

\bibitem{BorosLiangMeng}
C.Boros, Liang Zuo-tang, and Meng Ta-chung, 
Phys.Rev.Lett.{\bf 70} 1751 (1993);
C.Boros, Liang Zuo-tang, and Meng Ta-chung, 
Phys.Rev.{\bf D54} 4680 (1996);

%

\bibitem{Donnelly}
T.W.Donnelly, in Perspectives in Nuclear Physics at intermediate 
energies, eds. S.Boffi, C.Ciofi degli Atti, and M.M.Giannini (World 
Scisntific, Singapore, 1983), p.305. 
T.W.Donnelly and S.Raskin, Ann.Phys. (NY) {\bf 169} 247 (1986); 
T.W.Donnelly and S.Raskin, Ann.Phys. (NY) {\bf 191} 78 (1989); 

\bibitem{BB95}
A.Bianconi and S.Boffi, Phys.Lett..{\bf B348}, 7 (1995); 

\bibitem{BR97}
A.Bianconi and M.Radici, Phys.Rev.{\bf C56}, 1002 (1997); 

\bibitem{BGPR}
S.Boffi, C.Giusti, and F.Pacati, Phys.Rep.{\bf 226} 1 (1993). 
S.Boffi, C.Giusti, F.Pacati, and M.Radici, ``Electromagnetic response 
of atomic nuclei'', Vol 20 of Oxford Studies in Nuclear Physics 
(Oxford University Press, Oxford, 1996). 

%

\bibitem{Collins93}
J.C.Collins, Nucl.Phys. {\bf B496}, 161 (1993); 


\bibitem{BrodskyEtAl02}
S.J.Brodsky, P.Hoyer, N.Marchal, S.Peigne, and F.Sannino, 
Phys.Rev. {\bf D 65} 114025 (2002).


\bibitem{BrodskyHwangSchmidt02}
S.J.Brodsky, D.S.Hwang, and I.Schmidt
Phys.Lett. {\bf B530}, 99 (2002); 
S.J.Brodsky, D.S.Hwang, and I.Schmidt
Nucl.Phys. {\bf B642}, 344 (2002); 

\bibitem{BoerBrodskyHwang03}
D.Boer, S.J.Brodsky, and D.S.Hwang, 
Phys.Rev. {\bf D 67}, 054003 (2003); 

\bibitem{Collins02}
J.C.Collins, Phys.Lett. {\bf B536}, 43 (2002); 

\bibitem{JiYuan02}
X.-D.Ji and F.Yuan, 
Phys.Lett. {\bf B543}, 66 (2002); 

\bibitem{BelitskyJiYuan03}
A.V.Belitsky, X.-D.Ji and F.Yuan, 
Nucl.Phys. {\bf B656}, 165 (2003); 

\bibitem{JiMaYuan}
X.Ji, J.P.Ma and F.Yuan, Phys.Rev. {\bf D 71} 034005
(2005); 
X.Ji, J.P.Ma and F.Yuan, Phys.Lett. {\bf B 597} 299 (2004). 


\bibitem{BBMP}
C.J.Bomhof, P.J.Mulders, and F.Pijlman, 
Phys.Lett.{\bf B 596}, 277 (2004). 
A.Bacchetta, C.J.Bomhof, P.J.Mulders, and F.Pijlman, 
Phys.Rev.{\bf D 72}, 034030 (2005). 
C.J.Bomhof, P.J.Mulders, and F.Pijlman, 
Eur.Phys.J.{\bf c 47}, 147 (2006). 

\bibitem{RatcliffeTeryaev07}
P.G.Ratcliffe and O.V.Teryaev, 
[arXiv:hep-ph/0703293]. 

\bibitem{CollinsQiu07}
J.Collins and J.-W.Qiu, Phys.Rev.{\bf D 75}, 114014 (2007). 

\bibitem{QiuVogelsangYuan07}
J.W.Qiu, W.Vogelsang, and F.Yuan, [arXiv:hep-ph/0704.1153],  
[arXiv:hep-ph/0706.1196]. 

\bibitem{Yuan03}
F.Yuan, Phys.Lett. 
{\bf B575}, 45 (2003).

\bibitem{GambergGoldsteinOganessyan03}
L.P.Gamberg, G.R.Goldstein, and K.A.Oganessyan 
Phys.Rev.
{\bf D67}, 071504(R) (2003).

\bibitem{BacchettaSchaeferYang04}
A.Bacchetta, A.Schaefer, and J.-J.Yang, 
Phys.Lett. {\bf B578}, 109 (2004).

\bibitem{LuMa04}
Z.Lu and B.-Q.Ma, 
Nucl.Phys. {\bf A741}, 200 (2004). 



\bibitem{Pobylitsa03} 
P.V.Pobylitsa, 
[arXiv:hep-ph/0301236 (2003)]. 

\bibitem{DalesioMurgia04}
U.D'Alesio and F.Murgia 
Phys.Rev. {\bf D70} 074009
(2004). 

\bibitem{Burkardt04}
M.Burkardt, Nucl.Phys. {\bf A735}, 185 (2004);
M.Burkardt and D.S.Hwang, Phys.Rev. {\bf D69}, 074032 (2004);
M.Burkardt, Phys.Rev. {\bf D69}, 057501 (2004).


\bibitem{Drago05}
A.Drago, Phys.Rev.
{\bf D71}, 057501 (2005). 

\bibitem{GoekeMeissnerMetzSchlegel06}
K. Goeke, S. Meissner, A. Metz, and M. Schlegel, 
Phys.Lett. {\bf B637} 241 (2006).

\bibitem{BoerVogelsang06}
D.Boer and W.Vogelsang
Phys.Rev. {\bf D74} 014004
(2006). 

\bibitem{Entropy3}
A.Bianconi, Phys.Rev. {\bf D 75} 074005 (2007). 

%

\bibitem{Torino05}
M.Anselmino et al, 
Phys.Rev.{\bf D72}, 094007 (2005); 

\bibitem{VogelsangYuan05}
W.Vogelsang and F.Yuan, 
Phys.Rev.{\bf D72}, 054028 (2005); 

\bibitem{CollinsGoeke05}
J.C. Collins, A.V. Efremov, K. Goeke, M. Grosse Perdekamp, 
S. Menzel, B. Meredith, A. Metz, and P.Schweitzer, 
Phys.Rev. {\bf D73} 094023 (2006). 

\bibitem{BR06a}
A.Bianconi and M.Radici, Phys.Rev. {\bf D 73} (2006) 034018. 

\bibitem{AnselminoDalesioMelisMurgia06}
M. Anselmino, U. D'Alesio, S. Melis, and F. Murgia, 
hep-ph/0608211. 

%

\bibitem{Star}
J.Adams et al (STAR), Phys.Rev.Lett. {\bf 92}, 171801 (2004);

\bibitem{Hermes}
A.Airapetian et al (HERMES), Phys.Rev.Lett. {\bf 94}, 012002 (2005);

\bibitem{Compass}
V.Y.Alexakhin et al (COMPASS), Phys.Rev.Lett. {\bf 94}, 202002 (2005);

\bibitem{Phenix}
S.S.Adler et al (PHENIX), Phys.Rev.Lett. {\bf 95}, 202001 (2005);


\bibitem{BaroneLuMa05} 
V.Barone, Z.Lu, and B.-Q. Ma
Phys.Lett. {\bf B632} 277 (2006). 


%

\bibitem{panda}
PANDA collaboration, 
L.o.I. for the {\sl Proton-Antiproton Darmstadt Experiment} (2004),  
{\tt http://www.gsi.de/documents/DOC-2004-Jan-115-1.pdf}.

\bibitem{assia}
GSI-ASSIA Technical Proposal, Spokeperson: R.Bertini, 
{\tt http://www.gsi.de/documents/DOC-2004-Jan-152-1.ps; }

\bibitem{pax} 
P.~Lenisa and F.~Rathmann [for the PAX collaboration],
hep-ex/0505054.

\bibitem{rhic2}
see e.g. G.Bunce, N.Saito, J.Soffer, and W.Vogelsang, Ann.Rev.Nucl.Part.Sci.
{\bf 50} 525 (2000) [arXiv:hep-ph/0007218]. 

\bibitem{compassDY}
R.Bertini, O.Yu Denisov et al, proposal in preparation. 

\bibitem{ShortLetter1}
A.Bianconi, [arXiv:hep-ph/0707.1240]

\bibitem{BR_JPG}
A.Bianconi and M.Radici, J.Phys.{\bf G 31} 645 (2005). 

\bibitem{Feshbach}
H.Feshbach, Ann.Rev.Phys. {\bf 5} 357 (1958).
H.Feshbach, Ann.Rev.Phys. {\bf 19} 287 (1962).
A.K.Kerman, H.McManus, and R.M. Thaler, Ann.Phys. {\bf 8} 551 (1959).

\bibitem{Krilov}
N.S.Krilov, ``Works on the foundations of statistical physics'',
Princeton University Press, Princeton, New Jersey, 1979. 
V.A.Fok and N.S.Krilov, JETP {\bf 17} 93 (1947). 
A.S.Davidov, ``Kvantovaja Mechanika'', Nauka, Moskow, 1981. 

\bibitem{JakobMuldersRodriguez97}
R.Jakob, P.J.Mulders and J.Rodriguez, Nucl.Phys.{\bf A626} 
937 (1997).








\end{thebibliography}
\end{document}